\documentclass[useAMS,usenatbib]{mnras}

\usepackage{mathtext,amssymb,amsmath}
\usepackage{epsfig}
\usepackage{graphics}
\usepackage{url}
\usepackage{adjustbox}
\usepackage{microtype}

\usepackage{times}

\usepackage[T1]{fontenc}

\newcommand{\be}{\begin{equation}}
\newcommand{\ee}{\end{equation}}
\newcommand{\beq}{\begin{eqnarray}}
\newcommand{\eeq}{\end{eqnarray}}

\def\ergcm2s{{\rm erg\,cm^{-2}\,s^{-1}}}

\title[Hard X-ray luminosity functions of CVs]{Hard X-ray luminosity functions of cataclysmic variables: \\ Joint {\it Swift}/BAT and  Gaia data}
\author[V. F. Suleimanov et al.]{Valery F. Suleimanov,$^{1,2,3}$\thanks{E-mail: suleimanov@astro.uni-tuebingen.de}
Victor Doroshenko,$^{1,3}$
and Klaus Werner$^{1}$
\\
$^{1}$Institut f\"{u}r Astronomie und Astrophysik, Kepler Center for Astro and
Particle Physics, Universit\"{a}t T\"{u}bingen, Sand 1, 72076 T\"{u}bingen,
Germany\\
$^2$Kazan (Volga region) Federal University,  Kremlevskaya str. 18, Kazan 420008, Russia\\
$^3$Space Research Institute, Russian Academy of Sciences, Profsoyuznaya 84/32,
117997 Moscow, Russia}

\pubyear{2022}

\begin{document}
\label{firstpage}
\pagerange{\pageref{firstpage}--\pageref{lastpage}}
\maketitle

\begin{abstract}
Cataclysmic variables  (CVs) are the most numerous population among the Galactic objects emitting in hard X-rays. Most probably, they are responsible for the extended hard X-ray emission of the Galactic ridge and the central Galactic regions.
Here we consider the sample of CVs detected in the all-sky hard X-ray {\it Swift}/BAT survey which were also detected by \textit{Gaia} and thus have reliable distance estimates. 
Using these data, we derive accurate estimates for local number density per solar mass ($\rho_{\rm M} = 1.37^{+0.3}_{-0.16}. \times 10^{-5} M_\odot^{-1}$)
and luminosity density per solar mass ($\rho_{\rm L} =  8.95^{+0.15}_{-0.1}\times 10^{26}$\,erg\,s$^{-1}$\,M$_\odot^{-1}$) for objects in the sample. These values appear to be in good agreement with the integrated Galactic ridge X-ray emission and Nuclear Stellar Cluster luminosities. Analysis of the differential luminosity
functions  $d\rho_{\rm M}/d(\log_{10} L_{\rm x})$ and $d\rho_{\rm L}/d(\log_{10} L_{\rm x})$ confirms that there are two populations of 
hard X-ray emitting CVs. Intermediate polars dominate at luminosities $L > 10^{33}$\,erg\,s$^{-1}$, whereas non-magnetic CVs and polars are much more
numerous but have lower luminosities on average. As a consequence, the contribution of these populations to the observed hard X-ray luminosity is almost equivalent. 
\end{abstract}



\section{Introduction}

Cataclysmic variables (CVs) are close binary systems 
with an accreting white dwarf (WD) as a primary star and, as rule as, a normal solar-like star as a secondary \citep{Warn:03}. 
The space number density of CVs is important for understanding of their evolution \citep[see e.g.][and references therein]{ZS20},
and was studied many times  (together with other their statistical properties) using various 
methods and energy bands. For instance, \cite{Pala.etal:20} estimated the number density  of the optically and ultraviolet (UV) selected CVs using a volume-limited  survey (within 150 pc from the Sun). However, it is difficult to estimate a CV luminosity density per unit mass using this approach, because of the necessarily small volumes allowed by sensitivities of the current surveys,  which implies that more rare luminous CVs could be missed.  Such objects are more easily identified in flux-limited surveys, especially all-sky surveys in X-rays, see e.g. \citet{Schwope.etal:02}, \citet{Sazonov.etal:06}, and  \citet{PK12}.
There the authors used the ROSAT all-sky survey in the soft X-ray band and the RXTE Slew Survey in the classical X-ray band 3-20 keV, which allowed to detect more remote objects and move far beyond the several tens of parsecs typical for volume-limited surveys. The most luminous CVs in this energy band are magnetic CVs, that is polars and intermediate polars (IPs), and thus such studies are highly complementary to studies in the optical band which mostly contain non-magnetic CVs. The main issue with this approach is the limited number of known objects with reliable distance estimates, especially at larger distances.

Here we study statistical properties of all known CVs emitting in hard X-rays, at $E > 15$\,keV in the flux-limited all-sky survey by \textit{Swift}/BAT. 
As already mentioned, hard X-ray surveys are best suited for detection of magnetic CVs. Indeed, in such objects the accretion flow is disrupted by the magnetic field of the WD and is confined by the magnetic field at least in the vicinity of WD surface. Arising hot, optically thin post-shock structures around magnetic poles with typical temperatures $\sim$\,10\,keV are sources of hard X-ray radiation
\citep[see e.g. reviews by][]{Mukai:17, deMartino.etal:20}. 

 On the other hand, non-magnetic disc CVs, dwarf novae (DNs), and nova-like systems,  are also sources of hard X-ray 
 radiation, for instance, the well studied nearby dwarf nova SS Cyg \citep{MPT04,Ishida.etal:09}. Their hard X-ray luminosities 
 are typically by factor of ten to hundred lower compared to magnetic CVs, because either only a tiny part of the liberated energy is released in the hard X-ray band (nova-like CVs and DNs during outbursts), or simply because of low accretion rates (DNs during quiescence).  Nevertheless, non-magnetic CVs also appear in hard X-ray surveys, and moreover, their total hard X-ray luminosity could be comparable with the total hard X-ray luminosity of IPs because they are much more numerous \citep{RK:03}.    

In general, luminosities of CVs in the hard X-ray band are not very high, and even the brightest IPs are typically fainter than $\sim$10$^{35}$\,erg\,s$^{-1}$. This means that we can only investigate in detail CVs situated relatively close to the Sun. At the same time, the total number of CVs (of all kinds) in the Galaxy is known to be very large and it was established that IPs are likely responsible for the unresolved extended hard X-ray emission of the Galactic ridge \citep{Krivonos.etal:07} and
the central region of the Milky Way \citep{Perez.etal:15}. Indeed, the unit stellar mass emissivity of the Galactic ridge in the hard 17-60\,keV energy band was estimated to (0.9-1.2)$\times 10^{27}$\,erg\,s$^{-1} M_\odot^{-1}$  \citep{Krivonos.etal:07}, which is close to value obtained by \citet{Revnivtsev.etal:08}
($1.3\pm0.3 \times 10^{27}$\,erg\,s$^{-1} M_\odot^{-1}$) for CVs from the analysis of INTEGRAL observations of 17 bright CVs emitting in hard X-rays.
The main problem of the analysis by \cite{Revnivtsev.etal:08} is the small size of the considered sample, which is mostly related to the fact that distances to CVs were poorly constrained for the majority of sources at that time. 
The study by \citet{PM14} using Swift/BAT observations of 17 bright IPs suffers from the same problems, that is low numbers of sources in the sample and poorly constrained distances for many of them.

Measurement of parallaxes for many CVs performed by
{\it Gaia} changed the situation qualitatively, however, up to now estimates of the luminosity and mass densities for hard X-ray emitting CVs were only carried out by \citet{Schwope18} using the same small number of sources. Here we address this shortcoming by using the recent 105-month \textit{Swift}/BAT catalogue of hard X-ray sources \citep{oh18}, which allowed to identify 79 CVs with well known distances from {\it Gaia} EDR3 and obtain significantly better constraints for mass densities and luminosity functions of hard X-ray emitting CVs presented in this work. 

Finally, we discuss the obtained updated constraints in light of the observed luminosities of the extended Galactic hard X-ray emission and estimated masses of the corresponding stellar structures of the Milky Way and find that they are in a good agreement, and conclude that the bulk of the observed hard X-ray emission from the Galactic ridge and bulge is indeed associated with CVs. 

\section{Source sample and data sources}
Considering that the most sensitive all-sky survey to date in the hard X-ray band is provided by \textit{Swift}/BAT, we limited our analysis to sources detected in the 105-months BAT catalogue\footnote{\url{https://swift.gsfc.nasa.gov/results/bs105mon/}}.
In this catalogue 75 sources were classified as CVs, and 6 sources were classified as Nova. We have also searched the 
unclassified sources in recent publications and added a few sources in the list, which brought up the total number of CVs detected by BAT to 87.
We have also verified that the 17 year \textit{INTEGRAL} Galactic survey \citep{Krivonos.etal:21} which is more sensitive in the Galactic plane does not contain any extra identified CVs. We note that this implies that CVs constitute $\sim6$\% of all identified BAT sources, so if one assumes same fraction for unidentified sources, one could expect $\sim7$ CVs among those. That is $\sim8$\% of all CVs in the sample and \textit{Swift}/BAT survey can be considered reasonably complete in context of CV identification.

Unfortunately, for some of the sources unambiguous identification of optical counterparts or distance estimates by {\it Gaia} are missing for various reasons, which reduces the total number of CVs in our sample to 79.
These objects are listed in Tables\,\ref{tab:int_ip} and \ref{tab:int_cv} and contain CVs both detected by BAT and in the recent catalogue of distances derived from Gaia EDR3 data by \cite{BJ21}.  The sample contains 51 IPs or very credible candidates to IPs, and 28 other CVs, including 14 polars and 2 magnetic CVs without more detailed classification. There are also 11 non-magnetic CVs and one object without any sub-class identification. The remaining eight CVs detected by BAT but lacking distance measurement are presented in Table\,\ref{tab:int_ucv}. 
We note also, that in the BAT catalogue one source, 1RXS~J122758.8-485343 (SWIFT J1227.8-4856), is listed as CV, but is actually a gamma-ray emitting  Low Mass X-ray Binary \citep{deMartino.etal:13}. This object was obviously omitted from our study. The luminosities of individual sources and their distances from the Galactic plane were calculated using the distances estimated by \cite{BJ21} based on the {\it Gaia} EDR3 \citep{eDR320} parallaxes. Full posterior probabilities for geometric distances reported by \cite{BJ21} were used, i.e. all estimates reported in this work include the uncertainties associated with the uncertainty in distance for individual objects as discussed below. 
We note that \cite{BJ21} also provide photo-geometric distance estimates which take into the account brightness and colour of objects around a given position and might be more appropriate for sources with poorly constrained parallaxes. We note, however, that CVs have intrinsically peculiar colors so photo-geometric priors might lead to biased distance estimates, so conservatively rely on a more simple solution. It is important to emphasize also that the choice of the priors is actually not important for most of the objects in our sample as those have well constrained parallaxes dominating any prior choice. 
Finally, it is important to note that we impose no additional cuts on quality of astrometric solutions (characterized e.g. by excess astrometric noise or renormalised unit weight error (RUWE) in Gaia EDR3) as all of the objects considered here are binaries and thus have formally poor astrometric solutions. We emphasize, however, that a formal low quality of solutions does not necessary invalidate resulting distance estimates. Indeed, as discussed by \cite{Stassun21}, there appears no statistically significant difference between distances to eclipsing binaries derived from Gaia parallaxes (with large RUWE values) and those determined by other means.

\begin{table*}
\caption{Observed and derived data for intermediate polars. 
 \label{tab:int_ip} 
 }
{\footnotesize
\begin{center} 
\begin{tabular}{l l c c c  c c r}
\hline
\hline
Name & Swift Name & $f_{14-195}$ & d & $L_{14-195}$, & z& $\delta M$, & Type$^a$ \\
 &  & $10^{-12}$\,erg\,cm$^{-2}$\,s$^{-1}$ & pc & $10^{32}$\,erg\,s$^{-1}$ & pc & 10$^7$\,M$_\odot$ &\\
\hline
IGR J00234+6141	&	SWIFT J0023.2+6142	&	$14.1(2.6)$	&	$1557_{-72}^{+104}$	&	$41_{-9}^{+10}$	&	$-4.4_{-1.9}^{+1.6}$	&	$47_{-11}^{+14}$ & IP\\
V709 Cas	&	SWIFT J0028.9+5917	&	$76.6(2.7)$	&	$725_{-9}^{+9}$	&	$48.2(2.1)$	&	$-22.0(5)$	&	$57.4_{-2.6}^{+2.7}$   &IP\\
1RXS J005528.0+461143	&	SWIFT J0055.4+4612	&	$18.5(2.2)$	&	$930_{-20}^{+17}$	&	$19.2(2.4)$	&	$-245(6)$	&	$21.2_{-2.9}^{+2.8}$   & IP\\
GK Per	&	SWIFT J0331.1+4355	&	$61.0(2.7)$	&	$430_{-6}^{+7}$	&	$13.5_{-0.7}^{+0.8}$	&	$-53.6_{-1.4}^{+1.3}$	&	$14.5(9)$   & IP\\
1RXS J045707.4+452751	&	SWIFT J0457.1+4528	&	$18.5(3.0)$	&	$1498_{-172}^{+185}$	&	$49_{-12}^{+16}$	&	$64(5)$	&	$58_{-16}^{+22}$   & IP\\
V1062 Tau	&	SWIFT J0502.4+2446	&	$25(4)$	&	$1222_{-72}^{+88}$	&	$45_{-8}^{+9}$	&	$-195_{-14}^{+12}$	&	$53_{-10}^{+11}$   & IP\\
TV Col	&	SWIFT J0529.2-3247	&	$59.3(2.4)$	&	$502.8_{-3.5}^{+4}$	&	$18.0(8)$	&	$-234.6(1.8)$	&	$19.7(9)$   & IP\\
TX Col	&	SWIFT J0543.2-4104	&	$12.3(1.6)$	&	$909_{-21}^{+18}$	&	$12.1_{-1.6}^{+1.7}$	&	$-429_{-11}^{+10}$	&	$12.9_{-1.8}^{+2.0}$   & IP\\
V405 Aur	&	SWIFT J0558.0+5352	&	$33.9(2.6)$	&	$658_{-8}^{+8}$	&	$17.6(1.4)$	&	$185.1(1.9)$	&	$19.3_{-1.8}^{+1.7}$   & IP\\
MU Cam	&	SWIFT J0625.1+7336	&	$15.6(1.8)$	&	$931_{-19}^{+23}$	&	$16.1_{-1.9}^{+2.0}$	&	$403_{-8}^{+9}$	&	$17.5_{-2.2}^{+2.5}$   & IP\\
1RXS J063631.9+353537	&	SWIFT J0636.6+3536	&	$13.4(2.7)$	&	$1961_{-170}^{+192}$	&	$61_{-15}^{+18}$	&	$454_{-34}^{+42}$	&	$75_{-19}^{+24}$   & IP\\
V418 Gem 	&	SWIFT J0704.4+2625	&	$7.2(3.1)$	&	$3053_{-577}^{+866}$	&	$78_{-39}^{+63}$	&	$785_{-130}^{+186}$	&	$99_{-52}^{+100}$   & IP\\
BG CMi	&	SWIFT J0731.5+0957	&	$23.6(2.5)$	&	$867_{-24}^{+26}$	&	$21.2_{-2.5}^{+2.7}$	&	$223(6)$	&	$23.4_{-2.8}^{+3.2}$   & IP\\
SWIFT J073237.6-133109	&	SWIFT J0732.5-1331	&	$28.0(2.9)$	&	$1748_{-104}^{+131}$	&	$102_{-15}^{+18}$	&	$108_{-5}^{+6}$	&	$134_{-22}^{+30}$   & IP\\
PQ Gem	&	SWIFT J0750.9+1439	&	$32.5(2.9)$	&	$734_{-14}^{+16}$	&	$20.9(2.0)$	&	$271(4)$	&	$23.2_{-2.2}^{+2.4}$   & IP\\
EI UMa	&	SWIFT J0838.0+4839	&	$29.1(2.6)$	&	$1121_{-33}^{+27}$	&	$43_{-4}^{+5}$	&	$702_{-19}^{+20}$	&	$51(6)$   & IP\\
USNO-B1.0 0414-00125587	&	SWIFT J0838.8-4832	&	$15.7(3.2)$	&	$1670_{-146}^{+145}$	&	$51_{-12}^{+15}$	&	$-104_{-12}^{+10}$	&	$61_{-16}^{+20}$   & IP\\
DO Dra	&	SWIFT J1142.7+7149	&	$17.6(2.1)$	&	$194.9_{-1.0}^{+1.1}$	&	$0.8(1)$	&	$157.6(7)$	&	$0.53(8)$   & IP\\
V1025 Cen	&	SWIFT J1238.1-3842	&	$9.1(2.4)$	&	$196.0_{-3.0}^{+3.2}$	&	$0.4(1)$	&	$100.5(1.2)$	&	$0.23(8)$   & IP\\
EX Hya	&	SWIFT J1252.3-2916	&	$26.3(2.5)$	&	$56.77_{-0.05}^{+0.05}$	&	$0.10(1)$	&	$52.17(3)$	&	$0.033(4)$   & IP\\
IGR J15094-6649	&	SWIFT J1509.4-6649	&	$24.4(3.0)$	&	$1091_{-23}^{+24}$	&	$35(4)$	&	$-123.6_{-3.5}^{+3.4}$	&	$40(6)$   & IP\\
NY Lup	&	SWIFT J1548.0-4529	&	$95.2(3.5)$	&	$1266_{-30}^{+35}$	&	$183_{-12}^{+13}$	&	$173_{-4}^{+5}$	&	$268_{-20}^{+25}$   & IP\\
IGR J16500-3307	&	SWIFT J1649.9-3307	&	$22.5(2.4)$	&	$1091_{-52}^{+55}$	&	$32_{-4}^{+5}$	&	$158_{-6}^{+7}$	&	$37_{-5}^{+6}$   & IP\\
1RXS J165443.5-191620	&	SWIFT J1654.7-1917	&	$24.8(2.7)$	&	$990_{-43}^{+33}$	&	$29(4)$	&	$274_{-10}^{+11}$	&	$33_{-4}^{+5}$   & IP\\
V2400 Oph	&	SWIFT J1712.7-2412	&	$47.5(2.5)$	&	$700_{-11}^{+10}$	&	$27.8(1.7)$	&	$125.4(1.6)$	&	$31.5_{-1.9}^{+2.0}$   & IP\\
CXOU J171935.8-410053	&	SWIFT J1719.6-4102	&	$40.1(2.5)$	&	$622_{-12}^{+13}$	&	$18.6(1.3)$	&	$-3.9_{-0.5}^{+0.4}$	&	$20.4_{-1.5}^{+1.6}$   & IP\\
1RXS J173021.5-055933	&	SWIFT J1730.4-0558	&	$68.7(3.2)$	&	$1938_{-138}^{+180}$	&	$311_{-47}^{+61}$	&	$520_{-36}^{+44}$	&	$525_{-100}^{+120}$   & IP\\
IGR J18173-2509	&	SWIFT J1817.4-2510	&	$14.6(2.3)$	&	$4690_{-1071}^{+1452}$	&	$373_{-182}^{+368}$	&	$-339_{-141}^{+100}$	&	$663_{-377}^{+668}$   & IP\\
V1223 Sgr	&	SWIFT J1855.0-3110	&	$128.5(2.6)$	&	$561_{-8}^{+8}$	&	$48.4(1.6)$	&	$-119.7_{-1.9}^{+1.8}$	&	$57.6(2.1)$   & IP\\
V2306 Cyg	&	SWIFT J1958.3+3233	&	$13.6(2.7)$	&	$1253_{-42}^{+51}$	&	$26_{-5}^{+6}$	&	$56.9_{-1.2}^{+1.3}$	&	$29(7)$   & IP\\
V2069 Cyg	&	SWIFT J2123.5+4217	&	$17.8(2.6)$	&	$1155_{-39}^{+41}$	&	$28(5)$	&	$-94(4)$	&	$32_{-5}^{+6}$   & IP\\
RX J2133.7+5107	&	SWIFT J2133.6+5105	&	$55.0(2.9)$	&	$1372_{-35}^{+39}$	&	$124_{-9}^{+10}$	&	$9.64_{-0.31}^{+0.30}$	&	$169_{-14}^{+15}$   & IP\\
FO AQR	&	SWIFT J2217.5-0812	&	$52.9(3.2)$	&	$532_{-9}^{+7}$	&	$17.8(1.2)$	&	$-381(6)$	&	$19.6_{-1.5}^{+1.4}$   & IP\\
AO Psc	&	SWIFT J2255.4-0309	&	$31.0(2.5)$	&	$462_{-5}^{+4}$	&	$7.9(7)$	&	$-350_{-4}^{+3.5}$	&	$8.1(7)$   & IP\\
1WGA J0503.8-2823	&	SWIFT J0503.7-2819	&	$2.9(2.5)$	&	$837_{-43}^{+60}$	&	$2.4(2.1)$	&	$-456_{-39}^{+34}$	&	$2.4_{-1.6}^{+2.2}$   & $^1$IP\\
1RXS J052523.2+241331	&	SWIFT J0525.6+2416	&	$21(4)$	&	$1817_{-166}^{+183}$	&	$84_{-20}^{+27}$	&	$-176_{-23}^{+19}$	&	$108_{-29}^{+40}$   & $^1$IP\\
SWIFT J0535.2+2830	&	SWIFT J0535.2+2830	&	$12(4)$	&	$2625_{-774}^{+1607}$	&	$97_{-61}^{+179}$	&	$-74_{-62}^{+34}$	&	$126_{-83}^{+338}$   & $   ^2$IP\\
2MASS J06141230+1704321	&	SWIFT J0614.0+1709	&	$7.8(2.8)$	&	$1631_{-192}^{+270}$	&	$25_{-9}^{+12}$	&	$18.80_{-0.28}^{+0.21}$	&	$28_{-11}^{+15}$   & $^1$IP\\
SWIFT J0927.7-6945	&	SWIFT J0927.7-6945	&	$8.5(1.9)$	&	$1206_{-36}^{+34}$	&	$14.7_{-3.4}^{+3.5}$	&	$-261(8)$	&	$16(4)$   & $   ^2$IP\\
1RXS J095750.4-420801	&	SWIFT J0958.0-4208	&	$7.0(2.1)$	&	$1425_{-79}^{+89}$	&	$17_{-5}^{+6}$	&	$268_{-14}^{+15}$	&	$18_{-6}^{+7}$   &$^4$IP\\
IGR J14257-6117	&	SWIFT J1424.8-6122	&	$12.5(3.2)$	&	$2082_{-336}^{+451}$	&	$64_{-26}^{+52}$	&	$0_{-6}^{+4}$	&	$79_{-35}^{+80}$   &$   ^3$IP\\
2MASS J17012815-4306123	&	SWIFT J1701.3-4304	&	$10.6(3.0)$	&	$995_{-38}^{+41}$	&	$13(4)$	&	$6.8(5)$	&	$13(4)$   & $   ^4$IP\\
SWIFT J2006.4+3645	&	SWIFT J2006.4+3645	&	$14.4(3.2)$	&	$4459_{-1030}^{+1426}$	&	$337_{-193}^{+538}$	&	$211_{-63}^{+116}$	&	$572_{-368}^{+938}$   &  $^{14}$IP\\
1RXS J211336.1+542226	&	SWIFT J2113.5+5422	&	$7.5(2.4)$	&	$612_{-44}^{+41}$	&	$3.3_{-1.1}^{+1.3}$	&	$63.1_{-2.9}^{+3.3}$	&	$3.0_{-1.2}^{+1.3}$   & $   ^2$IP\\
IGR J18308-1232	&	SWIFT J1830.8-1253	&	$17.1(3.5)$	&	$2218_{-432}^{+512}$	&	$99_{-41}^{+87}$	&	$-31_{-19}^{+12}$	&	$128_{-59}^{+151}$ &  $^{5,6}$IP\\
CXO J183219.3-084030	&	SWIFT J1832.5-0863	&	$19(4)$	&	$1481_{-419}^{+670}$	&	$48_{-30}^{+113}$	&	$24.0_{-1.2}^{+2.7}$	&	$57_{-37}^{+178}$   & $^{7,8}$ IP\\
YY Sex	&	SWIFT J1039.8-0509	&	$9.8(3.2)$	&	$382_{-22}^{+31}$	&	$1.7(6)$	&	$292_{-17}^{+20}$	&	$1.4_{-0.6}^{+0.7}$   & $^9$IP?\\
IGR J12123-5802	&	SWIFT J1212.3-5806	&	$13.3(2.7)$	&	$2736_{-338}^{+359}$	&	$118_{-34}^{+50}$	&	$231_{-26}^{+34}$	&	$160_{-53}^{+86}$   & $^{10,11}$IP?\\
1RXS J161637.2-495847	&	SWIFT J1617.5-4958	&	$22.0(2.7)$	&	$1543_{-127}^{+141}$	&	$63_{-12}^{+14}$	&	$30.6_{-0.7}^{+0.9}$	&	$76_{-15}^{+20}$   & $^{12}$IP?\\
V2487 Oph	&	SWIFT J1731.9-1915	&	$17.8(3.3)$	&	$6407_{-1634}^{+1646}$	&	$832_{-367}^{+644}$	&	$863_{-206}^{+269}$	&	$1444_{-629}^{+726}$   & IP?\\
IGR J18151-1052	&	SWIFT J1815.2-1089	&	$11.6(2.9)$	&	$4099_{-1683}^{+2655}$	&	$228_{-175}^{+526}$	&	$224_{-104}^{+167}$	&	$359_{-298}^{+1000}$   & $^{13,8}$IP\\
\hline
\end{tabular}
\end{center}
}
Notes: a) IP - intermediate polars, AM - polars (AM Her type), mCV - magnetic CVs, dCV - non-magnetic (disc) CVs. If there is no additional reference, the type of the source was taken from the catalogue of \citet{RK:03} or from \citet{Suleimanov.etal:19}. 
1) \citet{HT:15}, 2) \citet{Halpern.etal:18}, 3) \citet{Bernardini.etal:18},
4) \citet{Bernardini.etal:17}, 5) \citet{Bernardini.etal:12} 6) \citet{TH:13}, 7) \citet{Sugizaki.etal:00}, 8) \citet{Masetti.etal:13}, 9) \citet{WW03},
10) \citet{Masetti.etal:10}, 11) \citet{Bernardini.etal:13}, 12) \citet{Masetti.etal:06}, 13) \citet{Lutovinov.etal:12}, 14) \citet{Hare.etal:21}.
\end{table*}
\begin{table*}
\caption{Observed and derived data for other CV sub-classes. 
 \label{tab:int_cv} 
 }
{\footnotesize
\begin{center} 
\begin{tabular}{l l c c c  c c r}
\hline
\hline
Name & Swift Name & $f_{14-195}$ & d & $L_{14-195}$, & z& $\delta M$, & Type$^a$ \\
 &  & $10^{-12}$\,erg\,cm$^{-2}$\,s$^{-1}$ & pc & $10^{32}$\,erg\,s$^{-1}$ & pc & 10$^7$\,M$_\odot$ &\\
 \hline
IW Eri	&	SWIFT J0426.1-1945	&	$5.5(1.7)$	&	$366_{-18}^{+20}$	&	$0.88_{-0.28}^{+0.31}$	&	$-217_{-15}^{+13}$	&	$0.60_{-0.23}^{+0.28}$   &AM\\
Paloma	&	SWIFT J0524.9+4246	&	$9.9(2.9)$	&	$582_{-20}^{+28}$	&	$4.0(1.2)$	&	$61.6_{-1.8}^{+2.0}$	&	$3.8(1.3)$   &AM\\
BY Cam	&	SWIFT J0542.6+6051	&	$32.7(2.4)$	&	$264.5_{-1.7}^{+1.9}$	&	$2.74_{-0.21}^{+0.20}$	&	$92.9(5)$	&	$2.43_{-0.23}^{+0.22}$   &AM\\
1RXS J070648.8+032450	&	SWIFT J0706.8+0325	&	$7.8(3.5)$	&	$206_{-4}^{+4}$	&	$0.4(2)$	&	$39.1(4)$	&	$0.21_{-0.11}^{+0.14}$   &$^1$AM\\
V834 Cen	&	SWIFT J1409.2-4515	&	$7.4(1.7)$	&	$107.6_{-0.8}^{+0.9}$	&	$0.10(2)$	&	$49.3(2)$	&	$0.033_{-0.010}^{+0.011}$   &AM\\
1RXS J145341.1-552146	&	SWIFT J1453.4-5524	&	$16.7(2.5)$	&	$215.7_{-1.1}^{+1.8}$	&	$0.9(1)$	&	$33.43(8)$	&	$0.65_{-0.13}^{+0.12}$   &$^{10}$AM\\
V2301 Oph	&	SWIFT J1800.5+0808	&	$15.3(2.7)$	&	$122.0_{-0.9}^{+0.8}$	&	$0.27(5)$	&	$52.10_{-0.21}^{+0.23}$	&	$0.129_{-0.031}^{+0.032}$   &AM\\
AM Her	&	SWIFT J1816.1+4951	&	$49.0(2.1)$	&	$87.93_{-0.24}^{+0.24}$	&	$0.45(2)$	&	$59.1(1)$	&	$0.26(2)$   &AM\\
V1432 Aql	&	SWIFT J1940.3-1028	&	$48.6(2.6)$	&	$450_{-7}^{+7}$	&	$11.8(7)$	&	$-100.4(1.9)$	&	$12.4_{-0.8}^{+0.9}$   &AM\\
CD Ind	&	SWIFT J2116.0-5840	&	$6.0(2.6)$	&	$235.3_{-3.2}^{+4}$	&	$0.4(2)$	&	$-135.4_{-2.8}^{+2.7}$	&	$0.22_{-0.12}^{+0.13}$   &AM\\
1RXS J221832.8+192527	&	SWIFT J2218.4+1925	&	$8.9(3.2)$	&	$241_{-6}^{+6}$	&	$0.6(2)$	&	$-101.6_{-3.1}^{+3.0}$	&	$0.38_{-0.18}^{+0.19}$   &$^4$AM\\
SWIFT J2319.4+2619	&	SWIFT J2319.4+2619	&	$10.0(3.4)$	&	$513_{-20}^{+18}$	&	$3.1(1.1)$	&	$-252_{-11}^{+10}$	&	$2.8_{-1.1}^{+1.2}$   &$^5$AM\\
IGR J19552+0044	&	SWIFT J1955.2+0077	&	$15(4)$	&	$165.5_{-1.5}^{+1.9}$	&	$0.49_{-0.14}^{+0.13}$	&	$-19.0(4)$	&	$0.28_{-0.10}^{+0.11}$   &$^{6,7}$AM?\\
1RXS J234015.8+764207	&	SWIFT J2341.0+7645	&	$5.5(2.2)$	&	$552_{-15}^{+15}$	&	$2.0(8)$	&	$159(4)$	&	$1.7(8)$   &$^1$AM?\\
1RXS J071748.9-215306	&	SWIFT J0717.8-2156	&	$6.5(2.2)$	&	$2571_{-630}^{+1196}$	&	$51_{-23}^{+45}$	&	$-168_{-62}^{+38}$	&	$60_{-29}^{+63}$   &$   ^2$mCV\\
SWIFT J2237.2+6324	&	SWIFT J2237.2+6324	&	$5.4(1.7)$	&	$1085_{-149}^{+182}$	&	$7.5_{-2.9}^{+4}$	&	$106_{-12}^{+17}$	&	$7.6_{-3.1}^{+5}$   &$   ^2$mCV\\
TW Pic	&	SWIFT J0535.1-5801	&	$16.3(1.9)$	&	$422.1_{-2.2}^{+2.3}$	&	$3.5(4)$	&	$-207.7(1.3)$	&	$3.2(4)$   &dCV\\
AH Men	&	SWIFT J0610.6-8151	&	$5.5(2.4)$	&	$491.3_{-2.8}^{+2.8}$	&	$1.6(7)$	&	$-212.1(1.4)$	&	$1.3_{-0.6}^{+0.7}$   &dCV\\
V347 Pup	&	SWIFT J0612.2-4645	&	$6.8(2.3)$	&	$290.3_{-1.1}^{+1.2}$	&	$0.7(2)$	&	$-108.8(6)$	&	$0.44_{-0.18}^{+0.19}$   &dCV\\
1RXS J074616.8-161127	&	SWIFT J0746.3-1608	&	$16.3(2.5)$	&	$625_{-8}^{+10}$	&	$7.6(1.2)$	&	$69.0(7)$	&	$7.8_{-1.4}^{+1.3}$   &$^{3}$dCV\\
V426 Oph	&	SWIFT J1807.9+0549	&	$12.9(2.4)$	&	$190.0_{-0.7}^{+0.6}$	&	$0.6(1)$	&	$61.0(1)$	&	$0.33_{-0.08}^{+0.09}$   &dCV\\
V603 Aql	&	SWIFT J1848.4+0040	&	$8.7(2.5)$	&	$315.4_{-3.5}^{+3.0}$	&	$1.0(3)$	&	$24.68(4)$	&	$0.75_{-0.26}^{+0.27}$   &dCV\\
V1082 Sgr	&	SWIFT J1907.3-2050	&	$12.3(2.0)$	&	$644_{-7}^{+7}$	&	$6.1(1.0)$	&	$-122.2_{-1.8}^{+1.7}$	&	$6.1(1.1)$   &dCV\\
SS Cyg	&	SWIFT J2142.7+4337	&	$49.2(2.3)$	&	$112.35_{-0.35}^{+0.4}$	&	$0.74_{-0.04}^{+0.035}$	&	$6.90(5)$	&	$0.483_{-0.028}^{+0.032}$   &dCV\\
RU Peg	&	SWIFT J2214.0+1243	&	$9.7(2.4)$	&	$271.3_{-1.5}^{+1.5}$	&	$0.9(2)$	&	$-134.3(9)$	&	$0.58_{-0.18}^{+0.19}$   &dCV\\
V* HL CMa	&	SWIFT J0645.4-1651	&	$7.0(2.5)$	&	$293.5_{-2.1}^{+2.2}$	&	$0.71_{-0.25}^{+0.26}$	&	$-24.2_{-0.4}^{+0.35}$	&	$0.47_{-0.20}^{+0.21}$   &dCV\\
Z Cam	&	SWIFT J0825.2+7312	&	$6.7(2.0)$	&	$213.5_{-1.3}^{+1.2}$	&	$0.4(1)$	&	$136.3(6)$	&	$0.19_{-0.07}^{+0.08}$   &dCV\\
SWIFT J0623.9-0939	&	SWIFT J0623.9-0939	&	$10.9(3.0)$	&	$1592_{-45}^{+50}$	&	$33(9)$	&	$-264_{-8}^{+7}$	&	$38_{-11}^{+12}$   &$^{1,2}$CV\\
\hline

\end{tabular}
\end{center}
}
Notes: a) IP - intermediate polars, AM - polars (AM Her type), mCV - magnetic CVs, dCV - non-magnetic (disc) CVs. If there is no additional references, 
the type of the source was taken from the catalogue of \citet{RK:03}. 1) \citet{HT:15}, 2) \citet{Halpern.etal:18}, 3)  \citet{TH:13}, 
4) \citet{Bernardini.etal:14}, 5) \citet{Shafter.etal:08}, 6) \citet{Masetti.etal:10}, 7) \citet{Tovmassian.etal:17}, 8) \citet{Masetti.etal:13},
9) \citet{Masetti.etal:06}, 10) \citet{Potter.etal:10}.

\end{table*}

\begin{table*}
\caption{Observed  data for hard X-ray emitting CVs without {\it Gaia} distances. 
 \label{tab:int_ucv} 
 }
{\footnotesize
\begin{center} 
\begin{tabular}{l l c c c  c c r}
\hline
\hline
Name & Swift Name & $f_{14-195}$ & d & $L_{14-195}$, & z& $\delta M$, & Type$^a$ \\
 &  & $10^{-12}$\,erg\,cm$^{-2}$\,s$^{-1}$ & pc & $10^{32}$\,erg\,s$^{-1}$ & pc & 10$^7$\,M$_\odot$ &\\
\hline
XY Ari & SWIFT J0256.2+1925 & 30.5(2.2) & -& -& -& -& IP \\
1RXS J082033.6-280457 & SWIFT J0820.6-2805 & 12.4(2.5) & -& -& -& -& $^1$IP? \\
1RXS J093949.2-322620 & SWIFT J0939.7-3224 & 6.1(3.4) & -& -& -& -& $^1$IP? \\
SWIFT J1408.2-6113 & SWIFT J1408.2-6113 & 11.5(3.4) & -& -& -& -& $^2$IP? \\
1RXS J082623.5-703142 & SWIFT J0826.2-7033 & 10.0(1.8) & -& -& -& -& $^{3,4}$CV \\
SWIFT J074940.0-321536 & SWIFT J074940.0-321536 & 5.8(1.8) & -& -& -& -& $^b$CV \\
SWIFT J080040.2-431107 & SWIFT J080040.2-431107 & 5.8(1.9) & -& -& -& -& $^b$CV \\
1RXS J121324.5-601458 & SWIFT J1213.2-6020 & 11.3(1.5) & -& -& -& -& $^5$CV \\
\hline

\end{tabular}
\end{center}
}
Notes: a) IP - intermediate polars, AM - polars (AM Her type), mCV - magnetic CVs, dCV - non-magnetic (disc) CVs. If there is no additional references, 
the type of the source was taken from the catalogue of \cite{RK:03}. b) the sources are identified as CVs in the BAT catalogue, 
but there are no additional identification references.  1) \citet{HT:15}, 2) \citet{Tomsick.etal:16}, 3) \citet{Parisi.etal:12},
4) \citet{Malizia.etal:17}, 5) \citet{Molina.etal:21}
\end{table*}

\begin{figure} 
\begin{center}  
\includegraphics[width=  0.99\columnwidth]{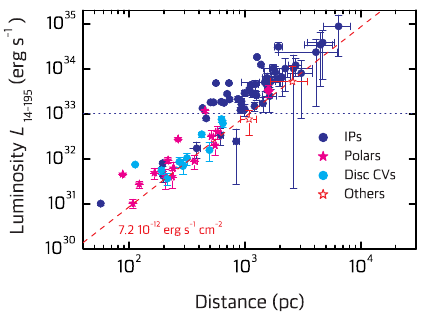}
\caption{\label{fig1} 
The luminosities of 79 observed CVs with known distances  vs. the distances to them. The distance-luminosity relation for the
adopted BAT sensitivity flux is also shown. 
The horizontal line at $10^{33}$\,erg\,s$^{-1}$ approximately separates IPs from other CVs. } 
\end{center} 
\end{figure}

\begin{figure} 
\begin{center}  
\includegraphics[width=  0.99\columnwidth]{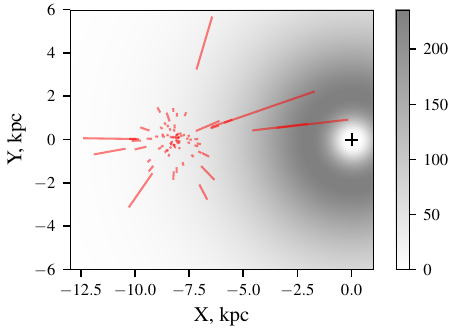}
\caption{\label{fig:gal_pl} 
Distribution of all the observed CVs in the Galactic plane. 
The positions of the sources with known distances together with the uncertainties are shown as red segments.
The surface density distribution in the thin disc in units M$_\odot$\,pc$^{-2}$ are shown as a grey scale. The center of the Milky Way is shown with the black cross. }
\end{center} 
\end{figure}

The strong apparent correlation of source luminosities and distance (see Fig.\,\ref{fig1}) clearly shows the flux-limited origin of \textit{Swift}/BAT survey, i.e. we only observe a tip of the iceberg of hard X-ray emitting CV population. This becomes obvious also when we inspect the location of the CVs in the sample within the Galaxy illustrated in Fig.~\ref{fig:gal_pl}, which shows that most of the objects in our sample are indeed relatively nearby. This is an expected consequence of relatively low CV luminosities and sensitivity of hard X-ray surveys.

Furthermore, it is well known 
\citep[see e.g.][]{Pretorius.etal:07} that most known CVs are members of the Galactic thin disk population, which is also a consequence of limited distance up to which these objects can be detected and reliably identified (i.e. it is hard to detect and identify CVs in the Galactic bulge and thick disk). We demonstrate this using our sample below.

It is also important to note that most IPs have luminosities 
above $10^{33}$\,erg\,s$^{-1}$, whereas other hard X-ray radiating CVs, including polars, have lower luminosities. This allows to distinguish objects of various types by their observed luminosities. We can, therefore,  reasonably suggest that the sources with an ambiguous identification and luminosities above  $10^{33}$\,erg\,s$^{-1}$ namely IGR J12123-5802, 1RXS J161637.2-495847, V2487 Oph, 1RXS J071748.9-215306, and SWIFT J0623.9-0939 are likely IPs. Note that the difference in typical luminosities for different CV types also implies that they can be observed up to different distances, which needs to be accounted for properly when modeling overall CV population. 

We conclude, therefore, that despite rather uniform coverage of the sky by BAT (sensitivity varies only by factor of two), the observed population of CVs is actually shaped by the selection effects associated with their spatial distribution in the Galaxy and limited (and luminosity-dependent) part of it sampled by BAT as expected for a flux-limited survey. Therefore, detailed modeling taking into the account spatial distribution of CVs within the Galaxy is required to obtain meaningful constraints on overall properties of their population.

\section{Method}
To recover the intrinsic luminosity and mass distributions of CVs one needs to account for the selection
effects mentioned above. For instance, this can be done using the modified $1/V$ method 
\citep[see e.g.][]{Schmidt68, Felten76}. Considering that CVs are distributed highly inhomogeneously, 
in our case integration over the Galactic mass distribution is more important than a simple volume
\citep[see e.g.][]{GGS02}. Therefore, one needs to assume some spatial distribution for CVs within the Galaxy, and then 
global population properties of the population 
can be described with three space densities
\be
          \rho_{\rm M} = \sum \limits^{N}_{i=1} \frac{1}{\delta M_i};\,~~~~~\rho_{\rm L} = \sum \limits^{N}_{i=1}  \frac{L_i}{\delta M_i};
          ~~~~~~\rho_{\rm N} = \rho_0\sum \limits^{N}_{i=1}  \frac{1}{\delta M_i},
\ee
i.e. source mass number ($\rho_{\rm M}$) and luminosity number ($\rho_{\rm L}$) densities normalized to solar mass, and
source volume number density per cubic parsec, $\rho_{\rm N}$. Here $\rho_{0}$ is the stellar mass density in the vicinity of the Sun 
measured in M$_\odot$\,pc$^{-3}$. Corresponding mass and luminosity functions are then determined as differentials of the source number and luminosity densities:
\be
       \Phi_{\rm M} = \frac{d\rho_{\rm M}}{d(\log_{10} L_{\rm x})};~~~~  \Phi_{\rm L} = \frac{d\rho_{\rm L}}{d(\log_{10} L_{\rm x})}.
\ee  
In practice this can be done by calculating corresponding densities for several luminosity bins. 

The mass of the Galaxy inside the volume where a given source would be detectable can be obtained by volume integration of the Galactic stellar mass density up to the maximal distance $d_{\rm max}$ defined by estimated actual source luminosity and sensitivity of the survey in given direction:
\be
      \delta M = \int \limits_0^{2\pi} dl  \int \limits_{-1}^{1} d\cos b \int \limits_0^{d_{\rm max}(l,b)} \rho(z,R)\, r^2 ~dr.
\ee
Here $l$ and $b$ are Galactic coordinates, $r$ is the current distance from the Sun, $R$ is a current distance from the Galactic center,
and $z$ is a current distance above the Galactic plane (the choice of cylindrical coordinates is motivated by the geometry of the Milky Way). Here it is important to emphasize that optical and soft X-ray surveys are strongly affected by interstellar absorption in the Galactic plane, so it is not trivial to estimate $d_{\rm max}$ in this case as it requires knowledge of spatial distribution of the absorbing material. On the other hand, the situation is much simpler in the hard X-ray band where absorption can largely be ignored.

The current values of $z$ and $R$, which are needed to calculate mass density
at a given point can be determined as
\be
      z = r \sin b; ~~~~~R^2 = (r\cos b)^2+R_0^2-2R_0 (r\cos b) \cos l,
\ee
where $R_0=8.1$\,kpc is distance from the Sun to the Galactic center. Surprisingly, there appears to be no single commonly accepted model for stellar mass distribution within the Milky Way,  see e.g. models presented by \citet{Juric.etal:08, McMillan11, Cautun.etal:20, Pieres.etal:20}.
We (somewhat arbitrarily) chose the relatively recent model described by \citet{Barros.etal:16}, where the mass distribution in all the Milky Way subsystems were presented. The latter detail is important considering that, as mentioned above,most of the known hard X-ray emitting CVs are members of the thin disc (see also the next Section), i.e. their density distribution at least in the vicinity of the Sun is expected to follow the mass distribution of this component only and one needs to be able to single it out. 
The stellar density of the thin disk can be described as double exponential along R- and z-coordinates \citep{Barros.etal:16}
\be
     \rho(z, R) = \frac{\Sigma_0}{2h}\exp\left[{-\frac{R-R_0}{R_d}-R_{\rm ch}\left(\frac{1}{R}-
     \frac{1}{R_0}\right)
     -\frac{\mid z\mid}{h}}\right],
\ee
where the local (near the Sun) surface density is 
$\Sigma_0 = 30.2$\,M$_\odot$\,pc$^{-2}$, 
he half-thickness of the disc is $h = 205$\,pc, 
nd the typical disc radius $R_d = 2.12$\,kpc.
The typical radius of the central hole in the thin disc $R_{\rm ch}$ is 2.07\,kpc. As discussed above, the observed vertical scale for hard X-ray emitting CVs is consistent with that of the stellar population, so we adopt the parameters by \cite{Barros.etal:16} as those are estimated based on the modeling of a significantly larger sample of normal stars, i.e. are better constrained.  

Once the spatial distribution is fixed, the values of $\delta M$ for each source can be calculated. The final values including the statistical uncertainties associated with the uncertainties in estimated distances and observed fluxes are presented in Tables\,\ref{tab:int_ip} and \ref{tab:int_cv}. To estimate statistical uncertainties related both to uncertainties in estimated flux/distance and limited number of sources in the sample, we used a bootstrap method. That is, we repeated the full analysis $10^{4}$ times, each time sampling the distances using the full posterior distributions published by \cite{BJ21}, and normally distributed fluxes from \cite{oh18} for each source. This allows to calculate $z$,  $L_{14-195}$,  $d_{max}$, and $\delta M$ for each realization (note that those are unambiguously defined), and then median value and $1\sigma$ confidence intervals for each source using the full sample (values reported in Tables\,\ref{tab:int_ip} and \ref{tab:int_cv}). The differential mass and luminosity density functions can also be calculated for each realization, and then uncertainties can be derived by averaging over all realizations. In the latter case, however, one needs to account also for the limited number of sources in each luminosity bin, which we do by adding associated Poisson uncertainty in quadrature to scatter over bootstrap realizations. The results are discussed in the next section.

We emphasize that the adopted stellar mass distribution model certainly does affect the derived integral numbers and luminosities for the CV population and thus there is some systematic uncertainty  in the derived parameters associated with choice of the stellar density model. 
We note, however, that this is a problem common for any population studies and in our case (i.e. with limited number of sources in the sample) is not a major concern as the estimated density model is derived using a much larger number of stars and describes their distribution adequately. 
An additional source of uncertainty is associated with the fact that \cite{BJ21} assume stellar density priors for object distribution whereas hard X-ray selected CVs tend to be more clustered to the Galactic plane. We note, however, that for most of the objects in our sample distance estimates are dominated by parallax uncertainties rather than density priors, so this is not expected to be a major issue.
A more serious concern is the uncertainty in the actual value of $d_{\rm max}(l,b)$ up to which source with a given luminosity $L_{14-195}$ can be detected, which is defined by the actual value of $L_{14-195}$ for a given source and the sensitivity of the survey in given direction. For objects in our sample most of the uncertainty in $d_{\rm max}$ is associated with uncertainty in distance and can be accounted for as described above. However, there is also a systematic uncertainty associated with actual sensitivity of the survey in a given direction. The distribution of estimated sensitivities of \textit{Swift}/BAT over the sky appears to be rather homogeneous ($\sim13$\%, see e.g. Fig.~10 in \citealt{oh18}), so we assumed a uniform limiting flux of $f_{\rm min} = 7.2\times 10^{-12}$\,erg\,s$^{-1}$\,cm$^{-2}$ to simplify the calculation of $\delta M$. This simplification is expected to introduce a systematic error of comparable magnitude (i.e. up to $\sim13$\%) in absolute values of the derived CV number and luminosity densities. We note that this systematic uncertainty is still smaller than the statistical uncertainty associated with the limited number of sources in the sample, but is still worth to keep in mind. 

\section{Results and discussion}

\subsection{Distribution over $z$-coordinate}

\begin{figure} 
\begin{center}  
\includegraphics[width=  0.99\columnwidth]{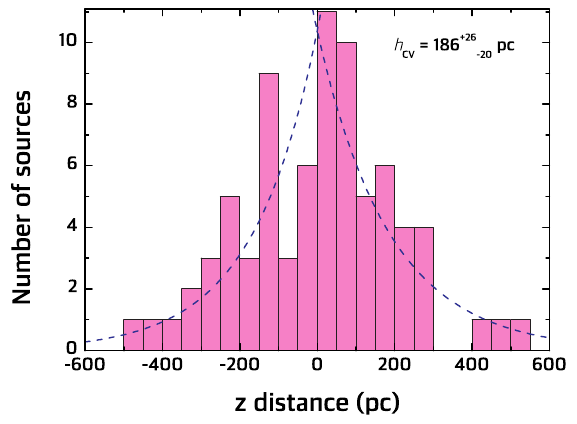}
\includegraphics[width=  0.99\columnwidth]{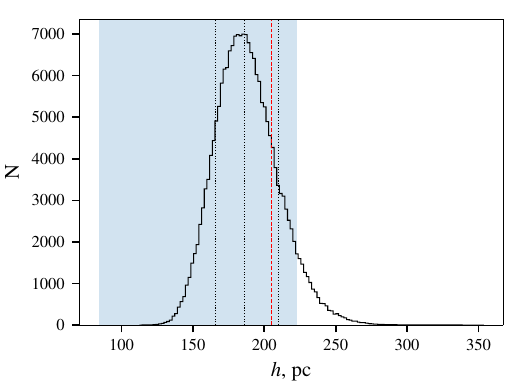}
\caption{\label{fig:z} 
Top panel: Distribution of all the observed CVs in z-direction. The best exponential fit is also shown.
Bottom panel: Posterior distribution of the exponential half-thickness $h$ for the sample of hard X-ray emitting CVs. The median value and one sigma confidence interval are indicated with dotted vertical lines. The adopted value of the stellar population thin disc half-thickness \citep[205 pc,][]{Barros.etal:16} is shown with the red dashed vertical line. The blue region depicts the range 
of $h_{\rm CV}$ estimated by \citet{Revnivtsev.etal:08}.} 
\end{center} 
\end{figure}

As discussed above, the stellar density within the thin disc can be approximated with an exponential distribution, and CVs are also expected to follow this. \cite{Revnivtsev.etal:08} approximated the spatial density of CVs as
$\rho_{\rm CV} \sim\exp{(-z/h_{\rm CV})}$ and derived a characteristic scale height of $h_{\rm CV}=130_{-46}^{+93}$\,pc. This is
  indeed comparable with the thickness of the Galactic thin disk estimated by \cite{Barros.etal:16} and other authors ($\sim200$\,pc). 
Our sample is significantly larger compared to that used by \cite{Revnivtsev.etal:08}, and spatial locations of individual objects are also much better constrained now. We repeated, therefore, the analysis by \cite{Revnivtsev.etal:08} using objects in our sample which allowed to estimate $h_{\rm CV}=186^{+26}_{-20}$\,pc. Here we use the same likelihood function as \cite{Revnivtsev.etal:08} and full distance priors to estimate the height above the Galactic plane for each source, and the \textit{emcee} package \citep{2013PASP..125..306F} to estimate the uncertainties for scale height from the posterior samples (presented in Fig.~\ref{fig:z}). The derived value is consistent within uncertainties with the  value of 205\,pc found by \cite{Barros.etal:16}, so our assumption that the observed CVs are members of the thin disk population is indeed justified and our modeling is self-consistent. Nevertheless, it is important to emphasize there are also some outlier sources with large $z$-distances, hardly compatible with being members of the thin disc, namely V2487 Oph ($z \approx$\,860 pc), V418 Gem ($z \approx$\,780 pc), and EI UMa ($z \approx$\,700 pc). Those objects might be, therefore, members of the thick disk, however, excluding them from the sample does not significantly affect our estimates, so we kept the thin disk assumption for simplicity.

\subsection{Number, mass, and luminosity density functions}
The number density of hard X-ray emitting CVs per solar mass  derived from the analysis described above 
is $\rho_{\rm M} = 1.37^{+0.3}_{-0.16} \times 10^{-5} M_\odot^{-1}$
and the corresponding luminosity density is $\rho_{\rm L} =  8.95^{+0.15}_{-0.1}\times 10^{26}$\,erg\,s$^{-1}$\,M$_\odot^{-1}$.
The corresponding volume number density of hard X-ray emitting CVs per cubic parsec is $\rho_{\rm N} = 8.6^{+1.8}_{-1.0} \times 10^{-7}$\,pc$^{-3}$.
Here the local stellar mass density near the Sun is assumed to be $\rho_\odot = \Sigma_0 /2h \times \exp{(-z_\odot/h)}\approx0.063$\,M$_\odot$\,pc$^{-3}$ where $z_\odot = 20.8$\,pc is the $z$-coordinate of the Sun \citep{Barros.etal:16}.

These values imply that the total number of hard X-ray emitting CVs is $\sim3.4\times10^5$ in the thin disk,  or $\sim5.9\times10^{5}$ in the Galaxy (thin and thick disks plus bulge) if a constant mass density $\rho_{\rm M}$ is assumed. Here the masses of the thin and thick disc components were assumed to be 2.5$\times 10^{10}$\,M$_\odot$ and 6$\times 10^{9}$\,M$_\odot$, respectively  \citep{Barros.etal:16}, and the bulge mass was taken as 1.3$\times 10^{10}$\,M$_\odot$  
\citep{Pieres.etal:20}.  
The corresponding integrated luminosities in the energy band 14-195\,keV are
$\sim2.2\times 10^{37}$\,erg\,s$^{-1}$ and 3.8$\times 10^{37}$\,erg\,s$^{-1}$ (for thin disk and all three components). 

 \begin{figure} 
\begin{center}  
\includegraphics[width=  0.99\columnwidth]{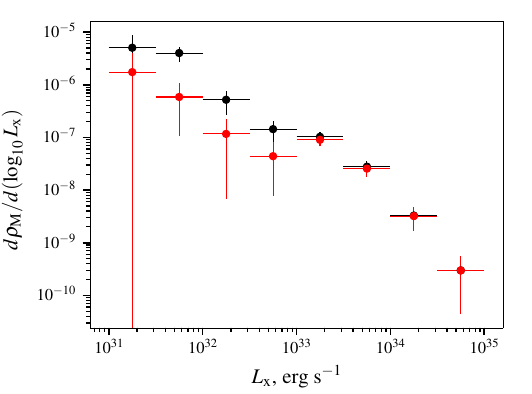}
\includegraphics[width=  0.99\columnwidth]{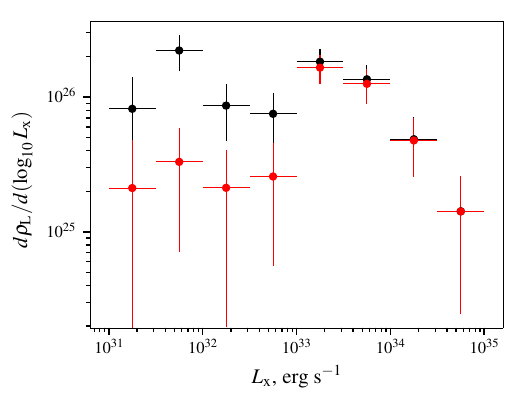}
\caption{\label{fig:lf} 
Top panel: Luminosity function of all the observed CVs,
demonstrating the number of CVs with different luminosities per solar mass.
Bottom panel: Differential luminosity distribution of 14-195 keV emissivity per
solar mass of all the observed  CVs. The contribution of the IPs 
is also shown (red circles).} 
\end{center} 
\end{figure}

The differential number density has, in general, a power-law shape index close to one (see upper panel in Fig.\,\ref{fig:lf}), although there are also clear deviations. These might be related to the diversity of the sources in our sample. To investigate this possibility, we also constructed luminosity functions separately for various types of CVs (see Fig.\,\ref{fig:lf}).  Indeed, it is evident that IPs dominate at luminosities above 10$^{33}$\,erg\,s$^{-1}$, but their contribution at lower luminosities 
is diminished and is by a factor of few less than that of other CV sub-classes (see also Fig.\,\ref{fig1}).  It means that the low-luminous non-magnetized
CVs are more common among hard X-ray emitting CVs. On the other hand, the contribution of IPs to the luminosity density is comparable 
with the contribution of other sub-classes because of their larger individual luminosities (see bottom panel in Fig.\,\ref{fig:lf}).  The difference between the two classes is most prominent in the luminosity function $\Phi_{\rm  L}$
which exhibits two corresponding peaks around  $\sim 6 \times 10^{31}$\,erg\,s$^{-1}$ and  $\sim 2 \times 10^{33}$\,erg\,s$^{-1}$.

\subsection{Comparison with previously published estimates}

 \begin{table*}
\caption{Comparison of the total space, mass, and luminosity densities of CVs. 
 \label{tab:int_lf} 
 }
{\footnotesize
\begin{center} 
\begin{tabular}[c]{ l | l l l l }
\hline
  &&&& \\
 Work   &  $\rho_{\rm N}$ & $\rho_{\rm L}$  &  $\rho_{\rm M}$   \\ 
  &(10$^{-7}$\,pc$^{-3}$)&(10$^{26}$\,erg s$^{-1}$ M$_\odot^{-1}$)&(10$^{-6}$\,M$_\odot^{-1}$)& \\
\hline
\hline
  &&&& \\
 This work	&  8.6$_{-1.0}^{+1.8}$  &     8.95$_{-0.1}^{+0.15}$   & 13.7$_{-1.6}^{+3.0}$  \\
 \citet{SDW20}$^a$ & $\approx$ 1.14  &  $\approx$ 8  & $\approx$ 3\\
 \citet{PM14}$^b$ &   1$^{+1}_{-0.5}$   &   & \\ 
\citet{Schwope18}$^c$ &	0.36$^{+0.4}_{-0.13}$&            &   \\ 
 \citet{Revnivtsev.etal:08}$^d$& 1.5$\pm$0.6 &  13$\pm$3 &  3.8$\pm$1.5 \\
 \citet{Sazonov.etal:06}$^e$&    & 24$\pm$6  &   12$\pm$3 \\
\citet{PK12}$^f$ & $40^{+60}_{-20}$      &    \\ 
 \citet{Pala.etal:20}$^g$ &  48$^{+6}_{-8}$  &  &    \\ 
 \hline
\end{tabular}
\end{center}
}
Notes: (a) 34 brightest IPs from 105-months BAT Catalogue were used;
(b) 15 IPs with  $L_{14-195} > 10^{32}$\,erg\,s$^{-1}$ from 70-months BAT Catalogue were used; (c) for the same IPs as in
\citet{PM14} but using {\it Gaia} DR2 distances and  assuming the height scale 200\,pc; (d) used 17 sources, 16 IPs and SS Cyg, observed with INTEGRAL;
 (e) using RXTE data in 3-20 keV energy band; 
(f) 20 non-magnetic CVs were used; (g) 43 optically and UV selected CVs, a volume-limited survey ($<$\,150\,pc). 
\end{table*}

Direct comparison of our results with some previously published estimates is presented in Table \ref{tab:int_lf}. 
The values of $\rho_{\rm N}$ obtained for hard X-ray emitting CVs using only bright IPs  by \citet{Revnivtsev.etal:08}, \citet{PM14}, \citet{Schwope18}, and \citet{SDW20} agree with each other and yield $\rho_N\sim10^{-7}$\,pc$^{-3}$ i.e. by factor of eight lower than what is obtained by us. There are several reason for this discrepancy. First,  $\rho_\odot\approx0.04$\,M$_\odot$\,pc$^{-3}$ estimated by \citet{Robin.etal:03} was used in these works, whereas we use a slightly larger value from more recent publication by \cite{Barros.etal:16}.
More important, however, is that in these works the low luminous
hard X-ray emitting CVs, such as polars and non-magnetic CVs, were not considered while in reality, as discussed above, they are dominant CV type at low luminosities. At the same time,  the estimations of $\rho_{\rm L}$ obtained by \citet{Revnivtsev.etal:08} and \citet{SDW20} are close to the 
value derived in this work. The reason is also obvious i.e. the prominence of IPs at high luminosities, which implies that their contribution to total luminosity is also dominant.

On the other hand, the values estimated by \citet{Sazonov.etal:06} are larger than what we derive likely because those are based on RXTE/PCA data in 3-20\,keV energy band, where the total luminosity is expected to be higher for all CV types, and furthermore, the contribution of non-magnetic CVs is significant. We also included the results for $\rho_{\rm N}$ obtained by \citet{PK12} for ROSAT selected CVs and derived by \citet{Pala.etal:20} for volume limited optically selected CVs. These results are close to each other regardless of large uncertainties obtained in the first work. It means that almost all CVs could be sources of the soft X-ray radiation. Comparison of the number densities obtained by \citet{Pala.etal:20} with our value demonstrates that only about twenty percent of all CVs emit in hard X-rays.

The general shapes of the luminosity functions  obtained by \citet{Revnivtsev.etal:08},  \citet{Sazonov.etal:06}, and \citet{SDW20}
are similar to what is derived in this work and presented in  Fig.\,\ref{fig:lf}. The functions $\Phi_{\rm M}$ are  also close to power-law with the exponents close to one,
and the functions  $\Phi_{\rm M}$ have their peak near luminosities $10^{32} - 10^{33}$\,erg\,s$^{-1}$, because the luminous IPs are well represented in these works. We note, however, that luminosity functions presented here have better sampling in luminosities \citep[including that presented in][] {SDW20}.

\subsection{Our results in broader astrophysical context}
As discussed above, CVs are among the most numerous population of sources emitting in hard X-rays. As a result, they have been suggested to account for unresolved hard X-ray emission coming from various parts of the Galaxy, which has no firmly established origin but would be hard to explain otherwise. 
Here we would like to discuss implications of the obtained results for several cases where hard X-ray emission of CVs might be relevant.

The first topic we would like to discuss is the so-called Galactic ridge X-ray emission (GRXE, \citealt{Krivonos.etal:07}). It was studied by the INTEGRAL observatory 
in the energy band 17-60 keV, and the authors of the cited paper found, that its integrated luminosity is $3.7\pm0.2 \times 10^{37}$\,erg\,s$^{-1}$.
This value is close to the value of derived by us above when contributions of thick disk and bulge are included, which we estimate to $3.94(6)\times 10^{37}$\,erg\,s$^{-1}$.  On the other hand, the  hard X-ray luminosity density necessary to explain the observed GRXE emission estimated by \cite{Krivonos.etal:07} is $(0.9 - 1.2) \times 10^{27}$\,erg\,s$^{-1}$ i.e. also close to the value obtained by us ($\rho_{\rm L} \approx 9 \times 10^{26}$\,erg\,s$^{-1}$).

Another interesting region where unresolved hard X-ray emission has been associated with CVs is the so-called Nuclear Star Cluster (NSC) observed near the Galactic center. It has a mass of about  $M_{\rm NSC} \approx 2.5 \times 10^7$\,M$_\odot$ and an effective radius about 5\,pc  \citep[see e.g.][and references within]{Nodueras.etal:21}. The size of the cluster is comparable to the size of the extended hard X-ray emission
at the Galactic center \citep{Perez.etal:15} discovered by the NuSTAR observatory. Recently \citet{Hailey.etal:16} suggested IPs as the likely source of observed extended hard X-ray emission from the NSC. The total luminosity of the NSC in the hard X-ray band is estimated to 2$\times 10^{34}$\,erg\,s$^{-1}$ \citep{Hailey.etal:16}. Scaling the value of
$\rho_{\rm L} =  8.95^{+0.15}_{-0.1}\times 10^{26}$\,erg\,s$^{-1}$\,M$_\odot^{-1}$ derived in this work to the total mass of the NSC we can estimate $\rho_{\rm L}\times M_{\rm NSC} \approx 2.24(4) \times  10^{34}$\,erg\,s$^{-1}$ i.e. very close to the observed value.
We can conclude, therefore, that the suggestion by \cite{Hailey.etal:16} that IPs are the dominant source of the observed emission is likely correct. That is, most of the observed emission comes from a few tens of IPs whereas total number of unresolved CVs is about $\rho_{\rm M}\times M_{\rm NSC} \approx 340$.

Finally, we can also compare our results with the recent observations and analysis performed by 
\citet{Bahramian.etal:20} for globular clusters (GCs). Taking an average  GC mass as $10^6$\,M$_\odot$ we expect their luminosities in hard X-rays to be about 
$10^{33}$\,erg\,s$^{-1}$, and estimate the total number of hard X-ray emitting CVs at about 10-15 per cluster. The number of luminous IPs must be 
very small, only 1-2 luminous IPs per globular cluster.  \citet{Bahramian.etal:20} investigated 38 GCs with a total mass 
$1.37 \times 10^{7}$\,M$_{\odot}$. We expect about 150-200 CVs emitting in hard X-rays, and about 40-50 luminous IPs in these GCs.
However, \citet{Bahramian.etal:20} have found only three luminous IPs in their sample. Most probably, this deficit is related to the age of the stellar population in GCs. Luminous IPs are relatively young objects and a large fraction of them can be expected to have already finished their evolution in GCs.
Alternatively, dynamical destruction could reduce the number of CVs in GCs \citep[][see, however, \citealt{Ivanova.etal:06}]{2019MNRAS.483..315B}.
The contribution of luminous IPs to the observed hard X-ray emission of the Galactic bulge can also be expected to reduced for the same reasons. \citet{Krivonos.etal:07} approximated the averaged spectrum of the Galactic ridge with a grid of  model IP spectra
\citep{SRR:05} and found that the average WD mass in IPs was about 0.5\,M$_\odot$. This value is lower than the average WD mass for a sample of nearby IPs, where it is about 0.8\,M$_\odot$ \citep[see e.g.][]{Suleimanov.etal:19, Shaw.etal:20}. The discrepancy between these values could be an indication that the contribution of IPs to the Galactic bulge is indeed reduced. On the other hand, the average WD mass in the NSC which was found from fitting of the spectrum of the central part of the Galaxy by  \citet{Hailey.etal:16} is about 0.9\,M$_\odot$, i.e. substantially higher than average value for the entire bulge. This could indicate that star formation processes may still be ongoing even in the vicinity of the Galactic center and thus the situation is more complex there than can be pictured by simple scaling of the derived CV densities to the mass of the bulge.

\section{Summary}
\label{sec:summary}
We presented here the luminosity functions and the number densities of hard X-ray emitting CVs based on the analysis of data from the all-sky 105-month BAT survey, and {\it Gaia} EDR3 distances \citep{eDR320, BJ21}. Combination of these two surveys allowed to compile an unprecedentedly large sample of 79 hard X-ray selected CVs with well constrained fluxes and Galactic positions. 
Such a large sample allowed us to significantly improve the existing constraints on number, mass, and luminosity densities of hard X-ray emitting CVs and corresponding luminosity functions for the local hard X-ray emitting CV population. 

We find integrated number and luminosity densities of
$\rho_{\rm M} = 1.37^{+0.3}_{-0.16} \times 10^{-5} M_\odot^{-1}$
and $\rho_{\rm L} =  8.95^{+0.15}_{-0.1}\times 10^{26}$\,erg\,s$^{-1}$\,M$_\odot^{-1}$ respectively.
Only the mass distributed in the thin disc of the Milky Way was taken into account in accordance with the model presented by 
\citet{Barros.etal:16} and the observed vertical scale of known CVs. The total thin disc mass in this model is about $2.5 \times 10^{10}$\,M$_{\odot}$, therefore, we expect about 340 thousands CVs emitting in hard X-ray in the thin disc with a total luminosity about 2.2$\times 10^{37}$\,erg\,s$^{-1}$.
Assuming that the derived number and luminosity densities are correct also for thick disk and bulge components, the numbers above can be scaled by factor of $\sim1.5$ to get corresponding estimates for the entire Milky Way.

We also demonstrated that the differential luminosity function  $\Phi_{\rm M} = d\rho_{\rm M} / d(\log_{10} L_{\rm x})$ has approximately a power-law
shape with an exponent close to one. The luminosity function also exhibits a feature at the luminosity $\sim 10^{33}$\,erg\,s$^{-1}$ where the total luminosity function starts to be dominated by emission from IPs. At lower luminosities the number density is defined by other sub-classes of CVs, 
which also dominate the total number density. Luminous IPs constitute about 26\% of the total CV population and yet contribute over half ($\sim52$\%) of the total luminosity in the hard X-ray band.
As a result,  the luminosity function  $\Phi_{\rm L} = d\rho_{\rm L}/d(\log_{10} L_{\rm x})$ exhibits two peaks associated with the two CV sub-types.

The obtained results are in a good agreement with the observed properties of the extended hard X-ray emission of the Galactic ridge 
\citep{Krivonos.etal:07} and for the Nuclear Star Cluster \citep{Perez.etal:15}. Their observed hard X-ray luminosities are perfectly reproduced with the luminosity functions obtained by us taking into account the masses of the NSC and the Galactic thin disc and the bulge. We also confirmed that the number density of luminous IPs in globular clusters is a few times smaller than in the vicinity of the Sun \citep[see, e.g.][]{Bahramian.etal:20} which is likely related to average age of the stellar population in GCs.

\section*{Acknowledgments} 
This work has made
use of data from the European Space Agency (ESA) mission {\it Gaia}
(\url{https://www.cosmos.esa.int/gaia}), processed by the {\it Gaia}
Data Processing and Analysis Consortium (DPAC,
\url{https://www.cosmos.esa.int/web/gaia/dpac/consortium}). Funding
for the DPAC has been provided by national institutions, in particular
the institutions participating in the {\it Gaia} Multilateral
Agreement.
The work was supported by the German Research Foundation (DFG) grant
WE 1312/53-1 (VFS),  and  Russian Science Foundation (grant 19-12-00423) (VFS and VD). VD thanks also 
the Deutsches Zentrum for Luft- und Raumfahrt (DLR) and DFG for financial support. 

\section*{Data availability}

The data underlying this article will be shared on reasonable request
to the corresponding author.



\bibliographystyle{mnras}
\bibliography{ip_lf} 

\end{document}